# Tuning electrochemical properties and thermal stability of YSZ mesoporous thin films for SOFC applications

B. Sievers, L. Granja, A. Zelcer, D. Juan., V. Ferrari, S. Passanante, V. Lombardo, M. C. Fuertes, R. Fuentes, J. Sacanell

**Abstract:**

Dense and mesoporous 4% mol Yttria-stabilized zirconia thin films were deposited on amorphous silica substrates. The effects of the temperature on the mesostructure were analyzed, by applying different thermal treatments to the samples within the range of SOFC operation conditions. Electrochemical impedance spectroscopy measurements rendered evidence of a single mechanism, consistent with grain boundary/surface oxide ion conduction. Our results show that the presence of the highly accessible porosity improves the superficial ionic conductivity, reducing the activation energy of the overall process. Density functional theory calculations on the (110) surface were performed to estimate the associated energy barriers. Our results suggest that ionic transport along the surface of accessible pores connected to the atmosphere could account for the observed reduction in activation energy.

**Introduction.**

Mesoporous oxides, synthesized through a combination of sol-gel processing method and surfactant self-assembly, exhibit exceptional characteristics including high surface area, controlled porosity, and an ordered pore structure adjustable within the 2–20 nm diameter range [Brinker1999]. Casting these materials as thin films unlocks a wide variety of applications, spanning from sensors to supercapacitors and biodevices [Innocenzi2013]. Notably, the high quality of mesoporous films, characterized by their minimal surface roughness, renders them highly suitable for integration into microdevice fabrication processes [Scalabenuzzi2023]. Moreover, the pores within these films are highly accessible and can be tailored to incorporate various species such as solvents, organic molecules, and nanoparticles, enabling precise control over material properties [Soler2011; Soler2012].

The prospective integration of mesoporous thin films (MTF) into contemporary material processing technologies holds significant promise. Their ability to be deposited on diverse substrates, combined with their high specific surface area and ordered 3-D accessible mesoporous structure, opens up numerous possibilities [Innocenzi2013]. For instance, tailored interfaces between different materials can be prepared from depositing films on various substrates or incorporating diverse materials within the porous structure, leading to alterations in surface electronic density. In such cases, the formation of a space charge region can be induced, significantly affecting the surface chemistry compared to bulk materials [Gregori2017], [Maler2009]. Furthermore, due to the crucial role played by interfacial interactions, the electronic and/or ionic conductivity of MTF are expected to be particularly enhanced.

Yttria-stabilized zirconia (YSZ) is considered the standard electrolyte material for Solid Oxide Fuel Cells (SOFC) [Jacobson2010] [Zacaria2019]. Its thermal, chemical and mechanical stability, in addition to its high ionic conductivity [will2000], make it an interesting functional material and a reliable component for SOFC and oxygen detectors [Singhal2002]. One of the main drawbacks of YSZ for SOFC implementation is the high operating temperature, typically around 1000°C. Such high temperatures impose strict requirements on the materials used for SOFC construction, as it leads to the degradation of the materials that make up the cell. Therefore, significant efforts are invested in developing both new materials and morphologies that can operate at lower temperatures [Zacarias2019]. Within this context, powder has been the most extensively explored morphology for ceramic and composite materials, where the presence of grain boundaries (GB) play a crucial role for ionic transport properties, especially at the nanoscale [Bellino2006], [Afroze2024]. Knöner et al. [Knöner2003] have demonstrated a significant improvement (up to three orders of magnitude) in oxide ion diffusion at grain boundaries compared to bulk materials in nanosized 6.9 mol% YSZ. These characteristics highlight the critical role of interfaces in the overall SOFC performance. In pursuit of reducing operating temperatures, recent research has been focused on exploring thin films geometry to minimize the path for oxide ions within the electrolyte while maximizing the interface between the electrolyte and the electrode. [CELIK2022].

In this work, we present the structural and electrochemical characterization of 4% mol YSZ mesoporous thin films, prepared using a commercial surfactant (Pluronic F127) as a templating agent. The use of a readily available reagent is an improvement with respect to previous reports that require custom made diblock polymers [Elm2015], [CELIK2022]. Structural stability of mesoporous oxides has been widely studied in the literature due to their sensitivity to high temperatures [Crepaldi2003], [Lionello2017], [Ha2012], [Hung2006]. One of the main challenges for MTF implementation consists in optimizing ionic transport properties preserving their structural stability within the working temperature range of SOFC [Zacarias2019]. With this aim, the samples were previously subjected to a protocol based on different thermal treatments (TT). The effects of the TT were thoroughly analyzed by X-rays reflectometry and diffraction. Through electrochemical impedance spectroscopy, we evaluated the in-plane ionic conductivity of the thermally treated films. By measuring within the operational temperature range, we determined the Activation Energy ($E_a$) of the overall electrochemical process. We also conducted studies on non-mesoporous YSZ thin films to understand the influence of the pores and their accessibility to the environment on the electrochemical properties.

In order to delve deeper into the mechanisms governing the ionic transport of YSZ mesoporous thin films, we performed ab initio calculations on the (110) YSZ surface. Specifically, Nudge Elastic Band (NEB) calculations were employed to explore oxygen diffusion pathways within the surface and into the bulk.

In summary, our theoretical-experimental studies unveil the potential of highly accessible mesoporous oxide thin films for SOFC technology and explore strategies to face their wear and tear due to the usual operation temperatures.

**Methods.**

*Experimental*

Mesoporous Yttria Stabilized Zirconia thin films were synthetized following the sol−gel route in combination with the evaporation-induced self-assembly strategy. [Brinker1999] A sol was prepared mixing a solution of Acac and Zirconium propoxide on absolute ethanol with a solution on $H_2O$, HCl and $YCl_3$ in ethanol. The molar ratio of Zr and of Y to total metal content (Zr+Y) was 0.92 and 0.08 respectively. A detailed procedure can be found in [Violi2015, Zelcer2013]. A templating agent, Pluronic F127, was added to the solution. This agent is used to produce the micelles which will become pores after a calcination procedure. The templating agent is used to produce the micelles, which will become pores after a calcination procedure. Films were deposited by dip-coating on fused silica substrates. After deposition, the mesostructure was stabilized with a thermal treatment of 2 hs at 60°C followed by a 2 hs at 130°C. Then the samples were heated up at 1°C /min to 350°C and calcinated during 2 hs, in order to obtain the porous structure. The same procedure but without the templating agent in the solution was followed to obtain dense (non-mesoporous) thin films.

Total reflection X-rays fluorescence (TXRF) measurements were performed to measure the actual Y and Zr ratio in the sol, using a Bruker S2 PICOFOX TXRF spectrometer. Sapphire sample discs were used to avoid the silicon signal. L1 lines of each element were used for quantification, and Ga was employed as an internal standard. Before performing the measurements, the sol was diluted using the following procedure: 200 µl of the sol were put in a 100,00 ml volumetric flask and the volume was completed with absolute ethanol. The concentration of Y and Zr present in the sol as obtained by TXRF are: [Y] = 0.039 ± 0.001 M and [Zr] = 0.41 ± 0.01 M.

In order to analyze the effects of temperature treatments on the films, we applied further thermal treatments. Dip coating allows the deposition of films with homogeneous thickness, pore structure and ordering over large areas (typically about 2 cm x 5 cm). This fact made possible to use the same porous and dense YSZ films, which were cut into pieces, to perform different thermal treatments and the subsequent experimental characterizations. Each sample was heated at a rate of 5°C/min up to a maximum temperature of 500°C, 650°C or 800°C, with a dwell time of 2 hours and then a free cooldown.

Grazing incidence X-ray diffraction (XRD) and X-ray reflectometry (XRR) measurements were performed using a Panalytical Empyrean diffractometer. The crystalline structure was determined by XRD. XRR was used to measure the thickness of the films. Their accessible porosity (AP) was determined comparing measurements performed under high and low relative humidity (RH) conditions (RH > 90% and RH < 15% respectively) [Fuertes2009] [Gibaud2006]. See Supplementary Information (SI) for more details.

The morphology was characterized by scanning electron microscopy (SEM) in a model SUPRA 40 Zeiss microscope.

To characterize the electrochemical properties, Pt microelectrodes 50 µm apart and 20 nm thick, were fabricated onto the films using conventional photolithography and Pt evaporation. Electrochemical impedance Spectroscopy (EIS) was measured between the microelectrodes in a probe station using a GAMRY 750 impedance analyzer. In order to determine the ionic conductivity parallel to the plane of the substrates as a function of temperature, impedance spectra were collected within a frequency range of 0.1 Hz – 100 KHz and a signal amplitude of 400 mV, at 450°C – 650°C in air atmosphere.

*Density functional theory and nudged elastic band calculations*

To complement the experimental results, we performed ab initio calculations using density functional theory (DFT) based on ultrasoft pseudopotentials (USPP) [Vanderbilt1990] as implemented in the Quantum Espresso (QE) code [Giannozzi2009]. Electronic exchange and correlation interaction was included within the generalized gradient approximation (GGA), by means of the PBE exchange functional. [PerdewPRL1996] Pseudo partial waves are generated in the Kleinman–Bylander fully non-local separable representation [Kleinman1982], [Rappe1990] and scalar relativistic approximation, available in the PS Library online database [QEPotentials], [DalCorso2014] (version 6.2.2 of the ATOMIC code). For the Zr (Y) element, $4s^2 4p^6 4d^2 5s^2$ ($4s^2 4p^6 4d^1 5s^2$) are considered as valence electrons. Lighter O atom is constructed with an electronic configuration that includes six valence electrons ($2s^2 2p^4$). In order to have tractable computational times but modelling as close as possible the experimental doping of YSZ, we constructed a 2x2x2 supercell (96 atoms) from the primitive unit cell. The doping was included by replacing four Zr atoms by four Y ones. For each pair of Y added, an oxygen vacancy was generated to maintain charge neutrality. The bulk doping yields a value of 6.3% mol. Different spatial distribution of Y atoms and oxygen vacancies were considered using two low-energy structures taken from a previous work [PARKES2015]. Brillouin zone integrations were carried out using a Monkhorst-Pack (MonkhorstPack1976) sampling of 2x2x2 points and an energy cutoff for wavefunctions (charge density) of 40 (400) Ry to attain converged results. Influence of the surface on ionic conductivity was studied by modelling slabs. Although the nanostructures could present different facets, in this study we focus on the (110) surface, for the sake of simplicity. The slab is constructed with a $2\sqrt{2}$x3 in-plane repetition of the unit cell keeping a 15 Angstrom separation between images to avoid spurious interactions. It has in total 106 atoms with the same absolute doping that represents a 5.6%mol concentration. The mesh of k-points was chosen to be 4x4x1. Nudged Elastic Band (NEB) [Henkelman2000] calculations were implemented between adjacent oxygen atoms along different directions to calculate the energy barriers involved in the ionic migration process. Lattice parameters and atomic positions were obtained by structural optimization, by allowing the structure to relax until the maximum component of the force acting on any atom was smaller than 0.001 Ry/au. In the slab structures, the atomic positions corresponding to the bottom plane were fixed to the bulk values.

**Results and discussion**

In Figure 1 we present the SEM micrographs of porous and dense samples (labeled as M-YSZ and D-YSZ, respectively), thermally treated at 350°C, 500°C, 650°C and 800°C (which will be referred to as M-YSZ-TT and D-YSZ-TT, where TT is the thermal treatment temperature). The inset of each micrograph shows the fourier transform (FT) of a section of each image of the M-YSZ samples. In the case of mesoporous samples (Figures 1.a to 1.d), the ordered porous structure is clearly observed in Fig. 1.a for M-YSZ-350 and is progressively lost while increasing the TT, a fact particularly clear for M-YSZ-800. For dense samples (Figures 1.e to 1.h), we can observe the grain growth induced by increasing TT. It is worth noting that both D-YSZ-800 and M-YSZ-800 present a very similar nanocrystalline morphology.

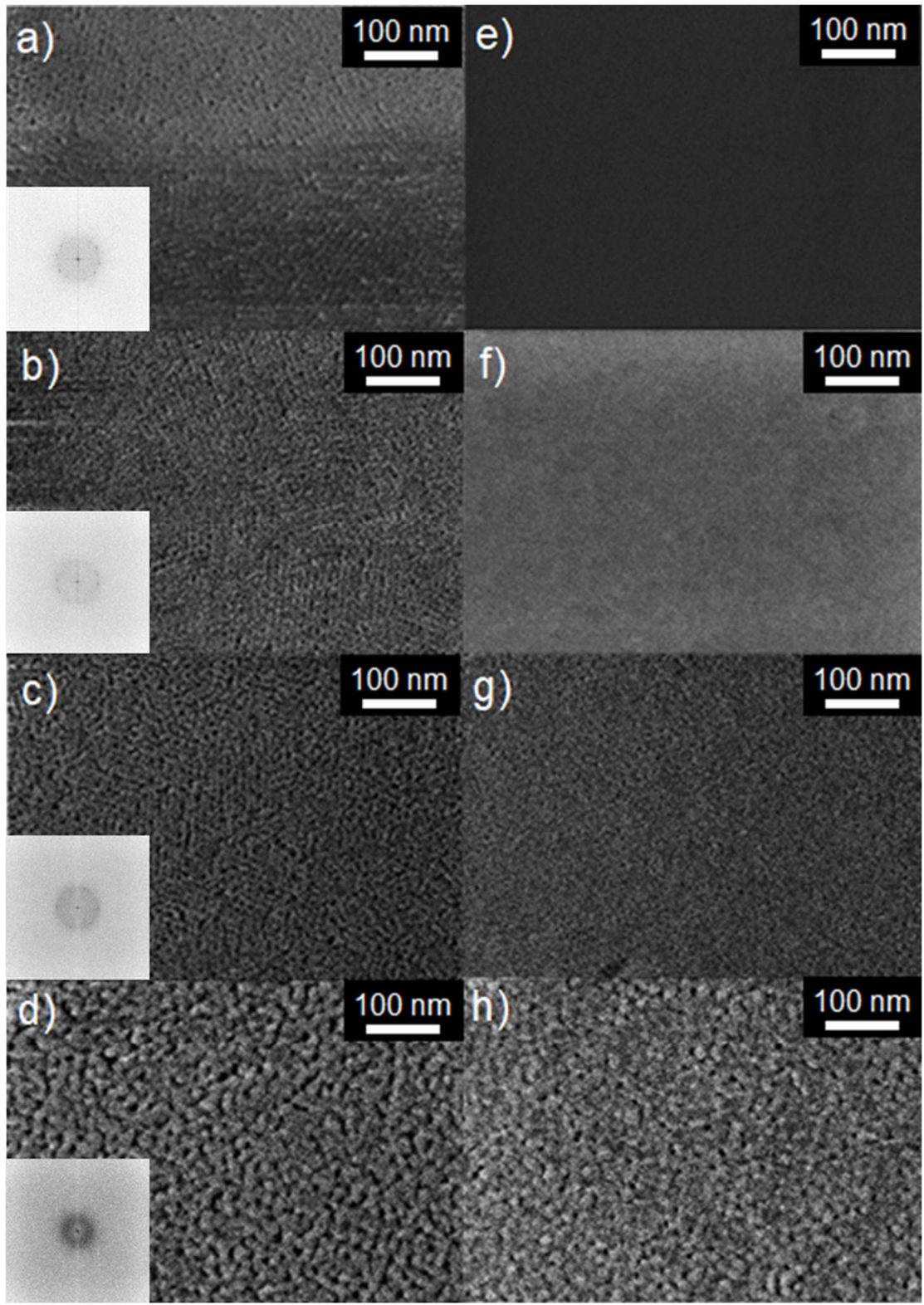

Figure 1: SEM images of the mesoporous (a, b, c and d) and dense (e, f, g and h) films after the corresponding thermal treatments: a) and e) 350 °C, b) and f) 500°C, c) and g) 650°C and d) and h) 800°C. The insets display the FT for each image of the M-YSZ samples.

X-ray diffraction patterns for M-YSZ and D-YSZ samples are presented in Figures 2.a and 2.b, respectively. The XRD patterns of the films treated at 350 °C shows no indication of crystallinity in the case of mesoporous films and just a weak reflection in the case of dense films. On the contrary, films which were thermally treated show many reflections, which become stronger with increasing TT. Depending on the temperature and the Yttrium content, bulk YSZ at atmospheric pressure can present monoclinic, tetragonal and cubic phases [Witz2007]. The observed peaks can be indexed either as cubic or tetragonal, but not as monoclinic, allowing us to rule out the presence of a significant amount of this phase. [Gotsch2016].

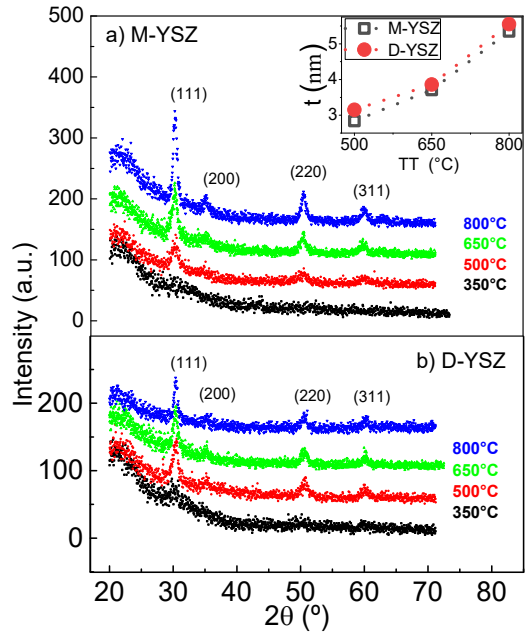

Figure 2: XRD patterns for samples with different maximum thermal treatment temperature: 350 °C, 500 °C, 650 °C and 800 °C, for a) M-YSZ and b) D-YSZ samples. Inset: Crystallite size obtained from the (111) peak vs. TT.

The inset of Figure 2.a shows the crystallite size ($t$) as a function of TT temperature, calculated with the Scherrer equation (Eq. S1, see SI), using the main XRD peak ((111) reflection). The crystallite size increases with TT temperature, showing similar values at each TT temperature for both D-YSZ and M-YSZ.

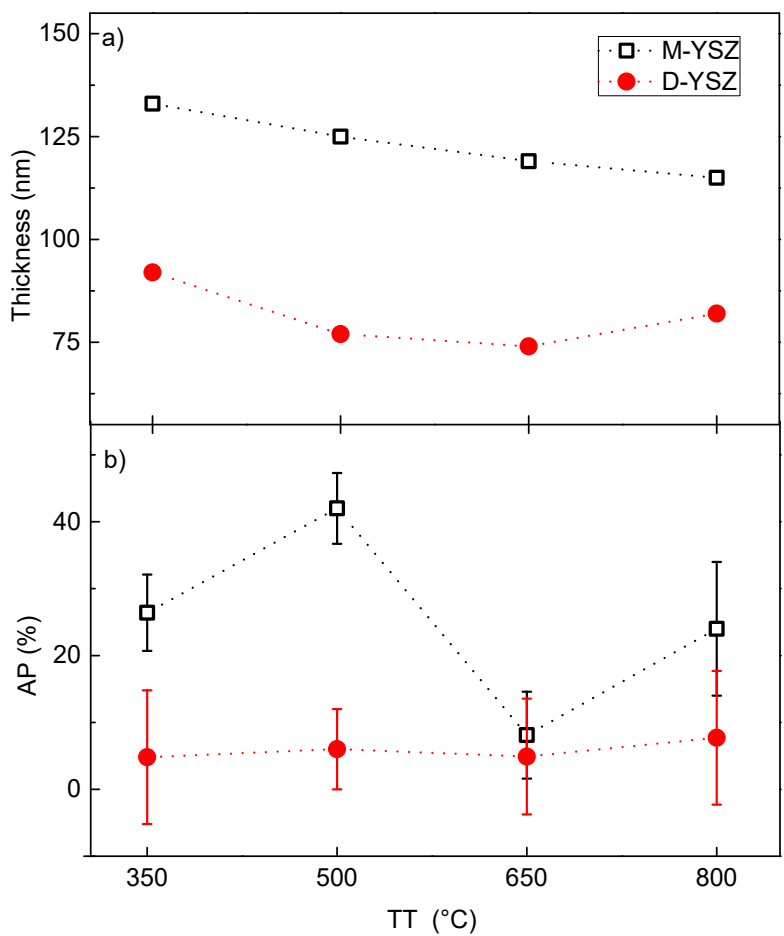

Figure 3: a) Thickness and b) accessible porosity, as a function of TT for M-YSZ (black, empty marks) and D-YSZ films (red, filled marks).

Figures 3.a and 3.b display the structural results (thickness and accessible porosity) obtained from the analysis of the XRR measurements (see Figures S1 and S2). The effect of thermal treatments on the films thickness is observed in Figure 3.a, showing that the behavior of mesoporous and dense films is different. On M-YSZ films, the increase of the TT temperature produces a monotonic reduction of the thickness of the film, suggesting that the increment of the grain size (see Fig. 1) affects the ordered mesostructure. The pores increasingly collapse with TT temperature, resulting in a systematic reduction of the thickness. Similarly, the thicknesses of D-YSZ films decrease with increasing temperatures at least up to 650°C, where a further increase of TT to 800°C leads to a slight increment of the thickness. This is most likely due to grain growth.

The accessible porosity is one of the most relevant properties of the mesoporous thin films. The AP to water vapor can be determined from XRR measurements performed in high and low humidity atmospheres (See Figures S.2b – S2.e) and Eq. S4 of the Supplementary Information). The AP results are shown in Fig. 3.b. In the case of D-YSZ films, AP remains lower than 10% for all thermal treatments, while M-YSZ films present AP ~ 30% - 40% for TT ≤ 500 °C. These results agree with the accessible porosity typically obtained for dense and mesoporous thin films, respectively. [Loizillon2019] However, for M-YSZ samples treated at 650 °C and 800 °C, AP shows a substantial decrease to ~10% becoming comparable to the AP of the D-YSZ films. This behavior is associated with the collapse of the porous structure, also observed in the SEM images where both samples treated at 800 °C have comparable morphology (Figures 1.d and 1.h).

In order to analyze the effect of the pores on the electrochemical properties, electrochemical impedance spectroscopy measurements were performed on M-YSZ and D-YSZ films at different temperatures, cycling from room temperature up to 650°C. The electrodes configuration is sketched at the inset of Figure 4. EIS measurements were performed cycling several times on the temperature range to evaluate the evolution of the samples while measuring: on heating from room temperature to the maximum measurement temperature (Ramp 1), on cooling to room temperature (Ramp 2) and on heating again (Ramp 3). SEM characterization shows that the mesoporous structure is preserved after this procedure. The Nyquist plots, obtained from EIS measurements of the samples with TT = 350°C and TT = 800°C are displayed in Figures 4.a to 4.d (See Figure S4 and S5 for a complete set of EIS and SEM images). Figures 5.a and 5.b compare the surface of sample M-YSZ-500 before and after these EIS measurements protocols.

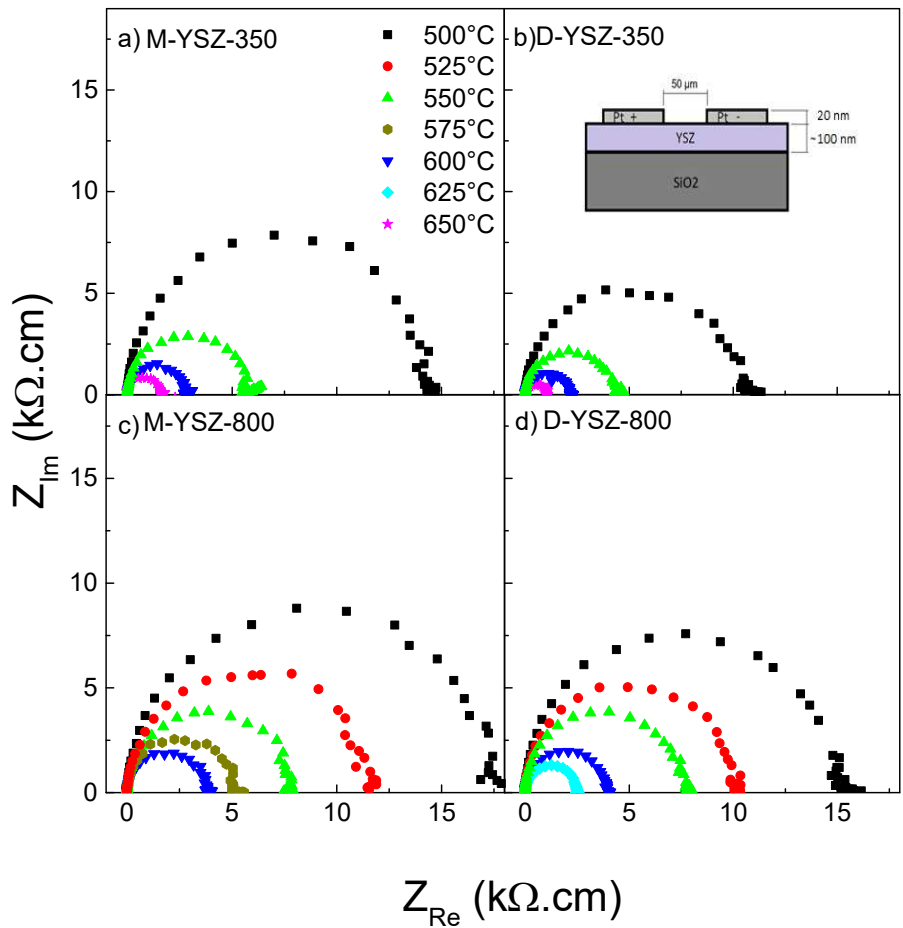

Figure 4: Electrochemical Impedance Spectroscopy spectra for M-YSZ (left) and D-YSZ (right) samples, a), b) TT at 350°C and c), d) TT at 800°C. In all cases, a single impedance arc (signature of a single dominant electrochemical process, associated with ionic conduction) is observed. Inset: schematic description of the Pt electrodes configuration for in-plane EIS measurement.

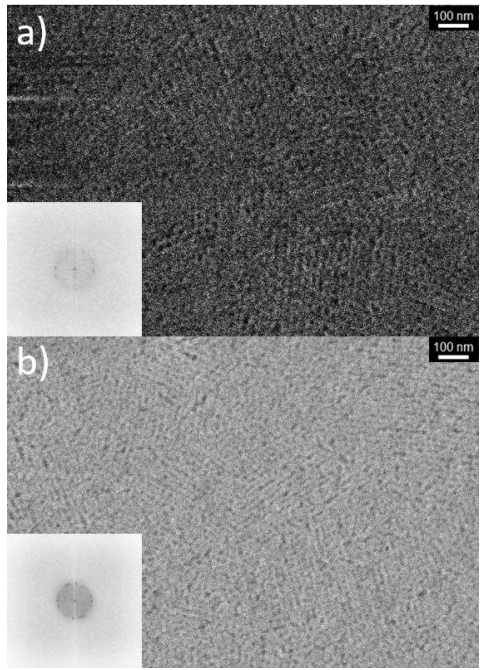

Figure 5: SEM the surface of sample M-YSZ-500 before (a) and after (b) the whole EIS measurements protocol.

The Nyquist plots of all samples present a single semicircle for each measurement temperature. Therefore, in all cases the electrochemical behavior is dominated by a single process that can be modelled as a parallel resistance-capacitor (RC) circuit, within the whole studied range. Particularly, the equivalent resistance (R) involved in the RC model is the difference between the low and high frequency intersects of the data with the real axis. Resistivity was calculated considering the distance between the electrodes and the cross section of the film, determined by the thickness resulted from each TT and the width of the electrodes. The ionic conductivity ($\sigma$) was obtained as the inverse of the resistivity.

Regarding the morphological changes induced by the TT performed to the samples, it is fundamental to discuss their implications in the ionic conductivity. Figure 6.a displays $\sigma$, calculated from the EIS data measured at 500°C, as a function of the TT temperature and Figure 6.b which depicts the ionic conductivity for M-YSZ samples measured at 500°C and 600°C, versus their accessible porosity. It can be observed that $\sigma$ presents a similar trend for D-YSZ and M-YSZ samples, being maximum for the samples with TT = 500°C. On one hand, the increase of $\sigma$ from TT = 350°C to 500°C is likely to be related to the crystallization, the formation of new grains and grain interfaces during the thermal treatment. On the other hand, the decrease of $\sigma$ for TT > 500°C, correlates with the increase of the grain size, which induces the collapse of the mesostructure and the consequent reduction of AP (Fig. 3.b). This is clearly visible in the data of Figure 6.b at 600°C. The correlation observed between $\sigma$ and AP highlights that the accessibility of the YSZ surface to the environmental oxygen favors the ionic conductivity.

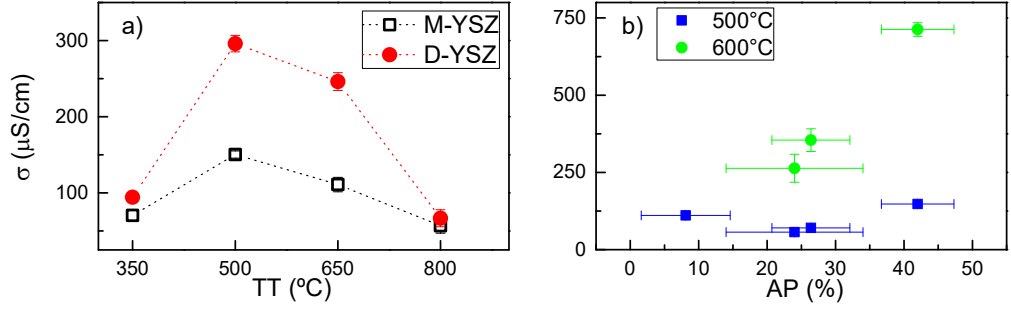

Figure 6: (a) Ionic conductivity of YSZ thin films measured at 500°C, as a function of the temperature of thermal treatment. (b) Ionic conductivity of M-YSZ vs the accessible porosity of the samples measured at 500°C and 600°C.

Figure 7 displays the Arrhenius Plot analysis of the ionic conductivity obtained from the EIS measured at different temperatures, for M-YSZ and D-YSZ films thermally treated at 500°C and 800°C. We can observe that Log($\sigma$) presents a linear behavior for the range of temperatures explored, suggesting that a single thermally activated process accounts for the conduction over all the range. Thus, it is possible to obtain the $E_a$ for each sample, by fitting the Arrhenius plot, according to eq. (1).

$$\sigma = A e^{-\frac{E_a}{k_B T}}, \qquad (1)$$

Where $A$ is a constant factor of the exponential and $k_B$ is the Boltzmann constant. The inset in Figure 6 shows the activation energy of the samples obtained. Activation energies range from 0.9 to 1.03 eV which are in good agreement with the 0.70- 1.2 eV values previously reported for YSZ. [Ikeda1985], [Chiodelli2005]. It is worth noting that the measured activation energy for M-YSZ is in all cases smaller than the one measured for D-YSZ for all TT.

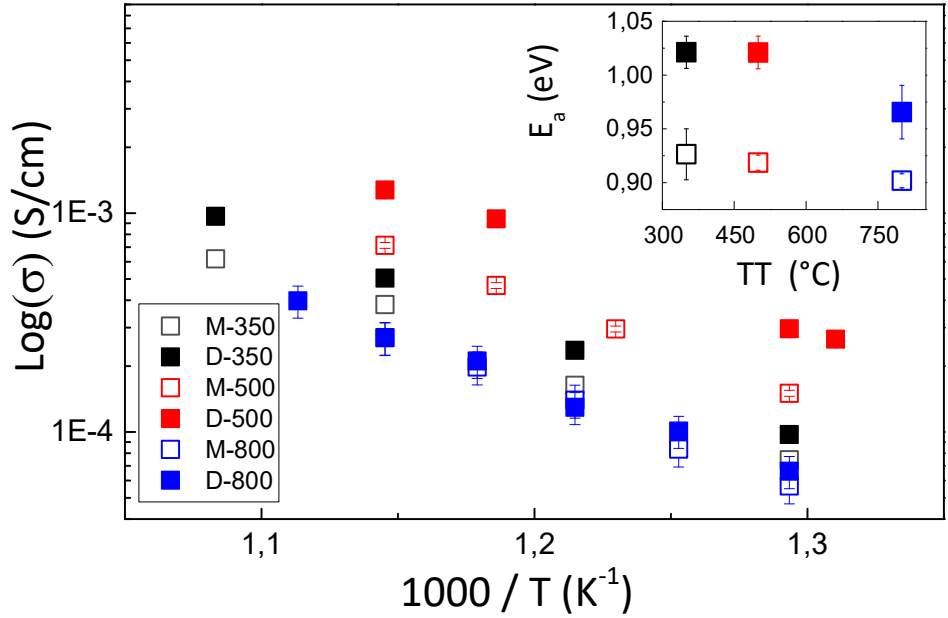

Figure 7: Arrhenius plot of the ionic conductivity for M-YSZ and D-YSZ thin films treated at 500°C and 800°C. Inset shows the activation energy vs temperature of thermal treatment obtained from eq. 6 while cooling to room temperature, see text for details.

In summary, both M-YSZ and D-YSZ systems have the same chemical composition and their crystallite sizes are comparable for each TT (inset of Figure 2), yielding a single mechanism in the EIS analysis of all these nanocrystalline YSZ thin films (Figure 4). Thus, the differences found in EIS for M-YSZ and D-YSZ systems should be understand in terms of the presence of the 3-D accesible mesoporosity (see Fig 3.d) and the evolution of the morphology with the thermal treatments. As expected, thermal treatments favor grain growth, decreasing the thickness, and destroying the mesoporous structure at the highest TT temperature studied (Figures 1- 3). However, we found that the M-YSZ sample with TT = 500°C has the optimum performance, preserving the structure even after several temperature cycles of EIS measurements (Figure 5). Within this framework, our findings demonstrate that the accessibly porosity favors the ionic conduction (Fig. 6.b) and that the mesostructure is crucial for diminishing $E_a$ (inset of Figure 7). Therefore, the evidence of a single mechanism for the ionic conduction, which is increased by the mesoporosity and its accessibility, strongly suggest that the ionic conduction develops mainly over the surface of the grains. This is supported by previous reports in nanoceramics, stating that grain boundary ionic conduction becomes the dominant electrochemical process with the reduction of the grain size [Bellino2006]. In this case, the AP would increase the surface exposed to the available environmental $O_2$, enhancing the superficial effect. We can then infer that the surface ionic conduction at the accessible pores is the main responsible for the lower $E_a$ of the M-YSZ samples (as compared to D-YSZ).

To gain deeper insight into the ionic transport mechanisms of nanostructured YSZ, we conducted Density Functional Theory (DFT) and Nudge Elastic Band (NEB) simulations focusing on the YSZ surface. Although both the (111) and (110) surfaces are the most stable ones, the (110) surface displays a higher vacancy concentration at the outermost oxygen layer [Iskandarov2015]. Therefore, in this paper we focus on the (110) surface as it should be the one primarily involved in the superficial ionic conduction.

For the slab simulation, we replicated the Yttrium atomic arrangement of the bulk structure. Specifically, we considered two different structures, named Slab1 and Slab2 (see Figure 8). The main distinction between those structures lies in the location of the Yttrium ions: in Slab1, they are located at the surface, while in Slab2, they are at the subsurface. These two slabs allowed us to explore the influence of the Yttrium environment in the oxygen transport.

It is noteworthy that the (110) surface exhibits an asymmetry within the plane concerning the O-O distance, with a smaller distance observed in the Y direction compared to the X direction. This will play a role in the oxide ionic transport, as we will show later. Further details regarding the calculations can be found in the Supplementary Information, including preliminary bulk analysis.

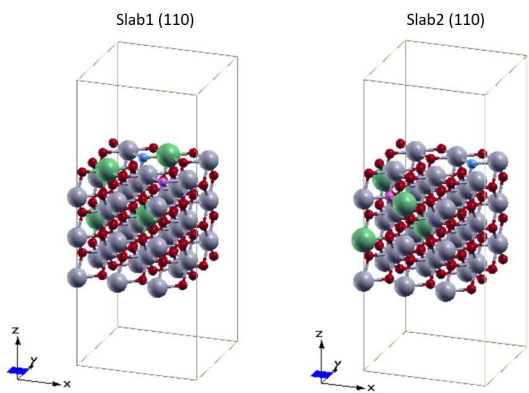

Figure 8: Supercells used for (110) slab calculations. Color-coding of the spheres correspond to: Yttrium (green), Zirconium (grey), Oxygen (red) atoms. Oxygen vacancies are depicted in magenta and blue. The blue vacancy is involved in the different NEB paths while the magenta one is fixed in all calculations.

For the NEB calculations, the initial configuration always presented an oxygen vacancy located at the surface, enabling exploration of various pathways, some of them within the surface (namely, X and Y directions) and others towards the bulk (with a Z component). We denote NEB paths as, for instance, $A^{\parallel}_{Slab2}$, where the letter labels the NEB path (A), the subscript indicates the slab, and the superscript indicates whether the NEB path is parallel to the surface ($\parallel$) or has a Z component ($\perp$) (refer to Table S2 for further details). The relative energy as a function of the relative coordinate (R) for both slabs are shown in Figure 9. From the NEB graphs we obtained the energy barriers, which are stated in Table 1. Due to the different Yttrium atomic environment, the NEB curves present an asymmetry. Therefore, we have to define a maximum and a minimun energy barrier ($E_{a\text{-}MAX}$ and $E_{a\text{-}MIN}$).

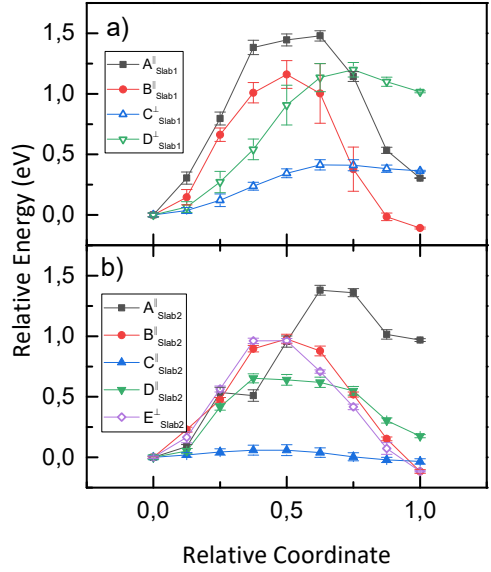

Figure 9: Energy profile for oxide ion migration along paths in the YSZ slabs. See Table S2 for path details, a) Slab1 and b) Slab2. Slab1 has a higher Y concentration closer to the surface compared to Slab2.

Table 1: Energy barriers extracted from Figure 9. The NEB label denotes the corresponding path, the subscript designates the slab number, and the superscript indicates whether the NEB path is parallel to the surface ($\parallel$) or if it has a Z component ($\perp$).

| NEB path | $E_{a\,MAX}$ (eV) | $E_{a\,MIN}$ (eV) |
|---|---|---|
| $A_{Slab1}^{\parallel}$ | 1.47 ± 0.04 | 1.17 ± 0.04 |
| $B_{Slab1}^{\parallel}$ | 1.20 ± 0.04 | 1.10 ± 0.04 |
| $C_{Slab1}^{\perp}$ | 0.41 ± 0.04 | 0.05 ± 0.04 |
| $D_{Slab1}^{\perp}$ | 1.17 ± 0.04 | 0.15 ± 0.03 |
| $A_{Slab2}^{\parallel}$ | 1.38 ± 0.04 | 0.41 ± 0.04 |
| $B_{Slab2}^{\parallel}$ | 1.09 ± 0.04 | 0.97 ± 0.04 |
| $C_{Slab2}^{\parallel}$ | 0.09 ± 0.04 | 0.06 ± 0.04 |
| $D_{Slab2}^{\parallel}$ | 0.65 ± 0.04 | 0.48 ± 0.04 |
| $E_{Slab2}^{\perp}$ | 1.08 ± 0.04 | 0.96 ± 0.04 |

As the Slab1 presents an Yttrium-rich surface, the endpoints of the $C_{Slab1}^{\perp}$ and $D_{Slab1}^{\perp}$ paths at R=0, have lower energies than the endpoint at R=1 (see figure 9). This indicates a preference for the oxygen vacancy to be located at the surface compared to the subsurface, by about 0.4 eV and 1.0 eV for $C_{Slab1}^{\perp}$ and $D_{Slab1}^{\perp}$ respectively. Regarding the NEB paths within the surface ($A_{Slab}^{\parallel}$ and $B_{Slab1}^{\parallel}$), there are differences in the energy barrier values due to the different atomic environment, as it will be discussed later. By contrast, the Slab2 structure displays mostly symmetric NEB paths due to its Yttrium-poor environment at the surface. In both structures, there is a tendency for the barriers in the Y direction to be lower than the barriers in X direction. This result is attributed to the smaller O-O distance along the Y axis present in the structure. In the Y direction, Slab2 presents barriers that are on average smaller or comparable to the mean bulk values (see Table S1). On the contrary, in the Z direction there is a significantly larger energy barrier (see Table 1). It is important to note that, using total energy calculations, we observe a general tendency for vacancies to be at the surface, which should favor surface conduction.

There is an increase in the energy barrier values for the case of Yttrium-rich concentration at the surface compared to the Yttrium-poor slab. This agrees with Molecular Dynamics calculations [Iskandarov2015] stating that higher local Yttrium concentration hinders oxygen migration. In line with that, Pornprasertsuk et al [Pornprasertsuk2005] found that vacancies tend to be at least next nearest neighbours to the Yttrium dopants, which in turn would affect the corresponding NEB barriers. Regarding oxygen migration away from the surface, we found that ion diffusion seems unlikely either due to the barrier being high or due to the mentioned preference for the vacancies to be at the surface. On the contrary, oxygen migration within the surface in Yttrium-poor conditions is favored due to the presence of smaller barriers in certain preferred directions. Our calculations suggest that within the myriad of possible paths involved, ionic conduction would mainly occur through those preferred paths that avoid high Yttrium concentration regions, as they tend to hinder ion conductivity. This scenario agrees with our experimental findings for mesoporous and dense YSZ thin films, as the former exhibits lower activation energies. This is likely attributed to the larger surface-to-volume ratio inherent in mesoporous materials. Furthermore, the experimental activation energy values (see inset of Figure 7) are within the range of energy barriers determined by our NEB calculations (Table 1).

To sum up, we can gather the above experimental and calculated results for a better understanding of the role of the mesoporous structure in the ionic transport properties. We can picture the M-YSZ system with the scheme shown in figure 10, displaying the nanocrystalline YSZ with pores. Many of the pores are connected to each other by open channels, typically named as necks, reaching up the film surface. Note that some necks can be closed because the neighboring YSZ grains constricting them and therefore rendering some of the pores inaccessible. This morphological feature is very important for SOFC applications which was quantified as the AP in this work. We conclude that the most energetically favorable routes for the oxide ions would be along segments of grain boundaries and pore wall surfaces. In order to illustrate this idea, we represented a possible route for the oxide ions in Fig 10. It is expected that macroscopic ionic current would develop in parallel over the myriad of all possible paths between the electrodes, with the contribution of the pore surfaces prevailing, which in turns reduces the activation energy of the system.

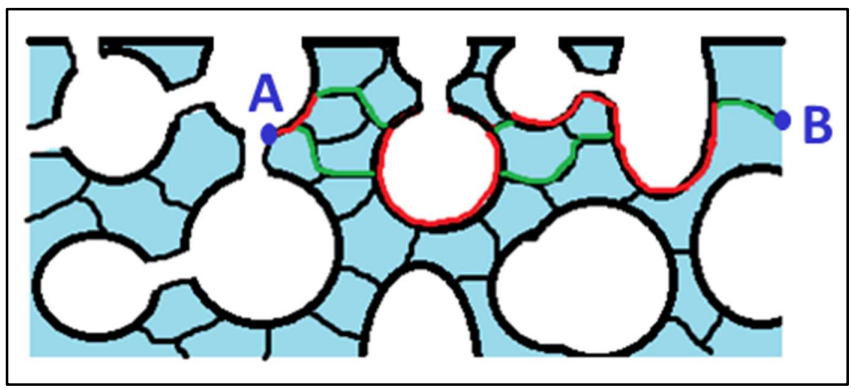

Figure 10: Diagram for the M-YSZ scenario. The schematic cross-section illustrates a possible path of ionic conduction from point A to point B, connecting grain-boundary paths (green) in series with pore surface paths (red).

**Conclusions**

In this work we developed mesoporous YSZ thin films, focusing on their potential as SOFC components. Based on the previously reported improvements of the ionic transport properties on nano-structuration [Bellino2006], [Abdala2010], we have chosen the prototypical YSZ oxide to look into the effects introduced by this morphology. Moreover, we used a commercial and widely applied templating agent to favor an earlier and easier implementation for SOFC applications.

We explored the best way to deal with the SOFC working temperature effects on the mesostructured material. Therefore, we prepared a set of mesoporous and dense YSZ thin films thermally treated for 2 hours at different temperatures (350°C, 500°C, 650°C and 800°C). We studied the evolution of the structure and the porous accessibility by XRD and XRR and their correlation with the ionic transport by EIS. The understanding of the experimental results was improved by NEB calculations of the oxygen vacancies in the YSZ surface under different structural conditions.

The NEB calculations suggest that the ionic conduction takes place through the more convenient paths, avoiding Yttrium-rich regions and mainly over the surface of the material instead of into the bulk. This is in agreement with our EIS experimental observations displaying a single dominating mechanism for ionic transport, which is mainly attributed to ionic conduction throughout the surface of the grains.

Our results demonstrate that the mesoporous thin films can bring important advantages to SOFC design. The strategies explored to preserve the mesostructure at the operative temperatures of the SOFC (approx. 700°C) found that the sample thermally treated at 500 ºC kept its structural and electrochemical properties through the thermal cycling protocol for EIS measurements. Furthermore, the tridimensional accessible porosity of the YSZ mesoporous thin films enhances the surface properties. This represents a promising advance for the integration of electrolyte and electrode materials in the development of SOFC based in thin films heterostructures.

**Credit authorship contribution statement**
Bernardo Sievers: Investigation, Formal Analysis, Writing- Reviewing and Editing
Iván Queriolo: Investigation, Formal analysis.
Verónica Lombardo, María C. Fuertes, Rodolfo Fuentes, Sebastián Passanante: Investigation, Methodology, Formal analysis, Writing-Reviewing and Editing

Dilson Juan., Valeria Ferrari: Investigation, Methodology, Software, Formal analysis, Writing-Original draft preparation, Writing-Reviewing and Editing
Andrés Zelcer: Investigation, Formal analysis, Methodology, Resources, Writing-Original draft preparation, Writing- Reviewing and Editing.
Leticia Granja and Joaquín Sacanell: Investigation, Formal analysis, Supervision, Conceptualization, Methodology, Resources, Writing-Original draft preparation, Writing-Reviewing and Editing, Project administration, Funding acquisition.

**Acknowledgements**

The present work was partially supported by Agencia Nacional de Promoción Científica y Tecnológica (Argentina, projects PICT 2017-858, PICT 2018-2397)